\def\a{1}  

\ifodd\a
\documentclass[a4paper,10pt,twocolumn]{article}
\usepackage{epsf}
\usepackage{amsbsy}
\usepackage{cite}
\else
\documentclass{article}
\usepackage{epsf}
\usepackage{amsbsy}
\usepackage{cite}
\fi

\newcommand{\ul}{\underline}

\begin{document} 

\date{\today}

\title{Modeling exchange bias microscopically}

\author{U.\ Nowak, A.\ Misra and K.\ D.\ Usadel\\
  Theoretische Physik, Gerhard-Mercator-Universit\"{a}t
  Duisburg,\\
 47048 Duisburg, Germany\\ e-mail: uli@thp.uni-duisburg.de}
\date{\today} 
\maketitle

\begin{abstract}
  Exchange bias is a horizontal shift of the hysteresis loop observed
  for a ferromagnetic layer in contact with an antiferromagnetic
  layer. Since exchange bias is related to the spin structure of the
  antiferromagnet, for its fundamental understanding a detailed
  knowledge of the physics of the antiferromagnetic layer is
  inevitable.  A model is investigated where domains are formed in the
  volume of the AFM stabilized by dilution.  These domains become
  frozen during the initial cooling procedure carrying a remanent net
  magnetization which causes and controls exchange bias. Varying the
  anisotropy of the antiferromagnet we find a nontrivial dependence of
  the exchange bias on the anisotropy of the antiferromagnet.
\end{abstract}



\section{Introduction}
\label{s:intro}

When a ferromagnet (FM) is in contact with an antiferromagnet (AFM) a
shift of the hysteresis loop along the magnetic field axis can occur
which is called exchange bias (EB).  Usually, this shift is observed
after cooling the entire system in an external magnetic field below
the N\'eel temperature $T_{\mathrm{N}}$ of the AFM.  Although this
effect is well known since many years\cite{meiklejohnPR56} and is
already intensively exploited in magnetic devices its microscopic
origin is still under debate. For a review of the experimental work
see the recent article by Nogu\'es and Schuller \cite{noguesJMMM99}.

In the approach of Malozemoff \cite{malozemoffPRB87} exchange bias is
attributed to the formation of domain walls in the AFM perpendicular
to the FM/AFM interface due to interface roughness. However, the
formation of domains in the AFM only due to interface roughness is
unlikely to occur because the creation of the domain walls is
energetically unfavorable.

Koon considered a spin-flop coupling between FM and the compensated
AFM as responsible for EB \cite{koonPRL97}, but recently, Schulthess
and Butler \cite{schulthessPRL98,schulthessJAP99} showed that
spin-flop coupling alone cannot account for this effect. Instead, in
their model EB is only obtained if uncompensated AFM spins are assumed
at the interface --- their occurrence is not explained
microscopically.

In a previous Letter Milt\'enyi et al.\ \cite{miltenyiPRL00} showed
that it is possible to strongly influence EB in Co/CoO bilayers by
diluting the antiferromagnetic CoO layer, i.\ e. by inserting
nonmagnetic substitutions (Co$_{1-x}$Mg$_x$O) or defects (Co\( _{1-y}
\)O) not at the FM/AFM interface, but rather throughout the volume
part of the AFM.  In these systems the observed EB is primarily not
due to disorder or defects at the interface. Rather, the full
antiferromagnetic layer must be involved and it was argued that in
these systems EB has its origin in a domain state in the volume part
of the AFM which triggers the spin arrangement and the FM/AFM exchange
interaction at the interface. This domain state develops due to the
dilution of the AFM: the domain walls pass preferentially through
non-magnetic sites thus reducing considerably the energy necessary to
create a wall.  The domain state strongly depends on the dilution of
the AFM resulting in a strong dependence of EB on dilution. Since
dilution favors the formation of domains it leads to an increase of
the magnetization in the AF and thus to a strong increase of the EB
upon dilution (see also \cite{mouginPRB01}, where it was shown that it
is possible to influence (to increase or even to revers) EB by a
subsequent ion irradiation of the sample).

In the same letter this picture was further supported by Monte Carlo
simulations. Later it was shown \cite{nowakJAP01,nowakPRB01} that a
variety of experimental facts associated with exchange bias can be
explained within our model, like positive exchange bias after cooling
in strong magnetic fields, the temperature dependence of exchange
bias, especially the relation between the so-called blocking
temperature and the N\'eel temperature, and the training effect, among
others. In these studies the AFM CoO investigated experimentally was
due to its rather strong uniaxial anisotropy modeled as Ising system
which is from a numerical point of view an ideal candidate to study
basic properties of EB. However, since the occurrence of EB is not
restricted to systems with a strong anisotropy in the AFM, in the
present paper we will extend the previous model
\cite{miltenyiPRL00,nowakJAP01,nowakPRB01} to the non-Ising case, i.\ 
e.\ we will vary the strength of the anisotropy of the AFM.

In the next section we give a brief review of the physics of domains
in diluted Ising antiferromagnets in an external field. These systems
have been studied in great detail in the past and the physics which
emerge from these studies are important for understanding EB.  In Sec.
\ref{s:model} our model is described and in Sec.  \ref{s:results} our
results from Monte Carlo simulations are discussed. Finally, we
summarize in the last section.

\section{Domains in disordered antiferromagnets} 
\label{s:domains}  

Considerable interest has been focused in recent years on the
understanding of {\bf d}iluted Ising {\bf a}nti{\bf f}erromagnets in
an external magnetic {\bf f}ield (DAFF) as they are ideal candidates
for the study of disordered systems. A typical material for
experimental studies is $\mathrm{Fe}_{1-p}\mathrm{Zn}_p \mathrm{F_2}$
where $\mathrm{Fe} \mathrm{F_2}$ is the AFM which is randomly diluted
with probability $p$ by non-magnetic Zn ions. Theoretically, due to
the very strong uniaxial anisotropy this system is usually treated as
Ising model. Properties which have been extensively exploited are the
critical behavior, domain structures, metastability and slow dynamics
(for reviews on DAFF see \cite{kleemannIJMP93,belangerBOOK98}).
Additionally, many of the findings of the DAFF are also relevant for
the {\bf r}andom {\bf f}ield {\bf I}sing {\bf m}odel (RFIM) which has
been shown to be in the same universality class
\cite{fishmanJPC79,cardyPRB84}.

In zero field the system undergoes a phase transition from the
paramagnetic phase to the long-range ordered, antiferromagnetic phase
at the disorder dependent N\'eel temperature $T_N$ as long as the
dilution $p$ is small enough so that the lattice of occupied sites is
above the percolation threshold. In the low temperature region, for
small magnetic fields $B$ the long-range ordered phase remains stable
in three dimensions \cite{imbriePRL84,bricmontPRL87}, while for higher
fields the DAFF develops a disordered domain state
\cite{montenegroPRB91,nowakPRB91} with a spin-glass-like behavior. The
reason for the domain formation was originally investigated by Imry
and Ma for the RFIM \cite{imryPRL75}.  The driving force for the
domain formation is a statistical imbalance of the number of
impurities of the two antiferromagnetic sublattices within a finite
region of the DAFF. This leads to a net magnetization of this region
which couples to the external field. A spin reversal of this region,
i. e. the creation of a domain can hence lower the energy of the
system. The necessary energy increase due to the formation of a domain
wall can be minimized if the domain wall passes preferentially through
non-magnetic defects at a minimum cost of exchange energy.  Hence,
these domains have non-trivial shapes following from an energy
optimization.  They have been shown to have a fractal structure with a
broad distribution of domain sizes and with scaling laws
quantitatively deviating from the original Imry-Ma assumptions
\cite{nowakPRB92,esserPRB97}.

In small fields the equilibrium phase of the three-dimensional DAFF is
long-range ordered. However, if cooled in a field $B$ below a certain
temperature $T_i(B)$, the system usually also develops metastable
domains \cite{birgeneauJSP84,belangerJAP85}.  The reason for this
metastability is a strong pinning which hinders domain wall motion.
These pinning effects are due to the dilution (random-bond pinning) as
well as due to the fact that a rough domain wall also carries
magnetization in a DAFF (following again the Imry-Ma argument) which
couples to the external field and hinders domain wall motion
(random-field pinning) \cite{villainPRL84}.  Consequently, after
cooling the system from the paramagnetic phase within an external
field, a DAFF freezes in a metastable domain state which survives even
after switching off the field, then leading to a remanent
magnetization which decays extremely slow \cite{hanPRB92}.

In the following we will argue that these well established properties
of the DAFF are the key for understanding exchange bias. Indeed,
during preparation of the system, the AFM is cooled in an external
magnetic field and additionally under the influence of an effective
interface exchange field stemming from the magnetized FM.  Hence, the
AFM --- as far as it is diluted in any sense --- must develop a domain
state with a surplus magnetization similar to that of a DAFF after
field cooling.

\section{Model for exchange bias} 
\label{s:model}  

The model which we consider in the following consists of one FM
monolayer exchange coupled to a diluted AFM film consisting of $t$
monolayers. In Fig.\ \ref{f:skizze} a sketch of our model is shown for
$t=3$.

\ifodd\a
\begin{figure}[h]
  \epsfxsize=55mm
  \hspace*{10mm} \epsffile{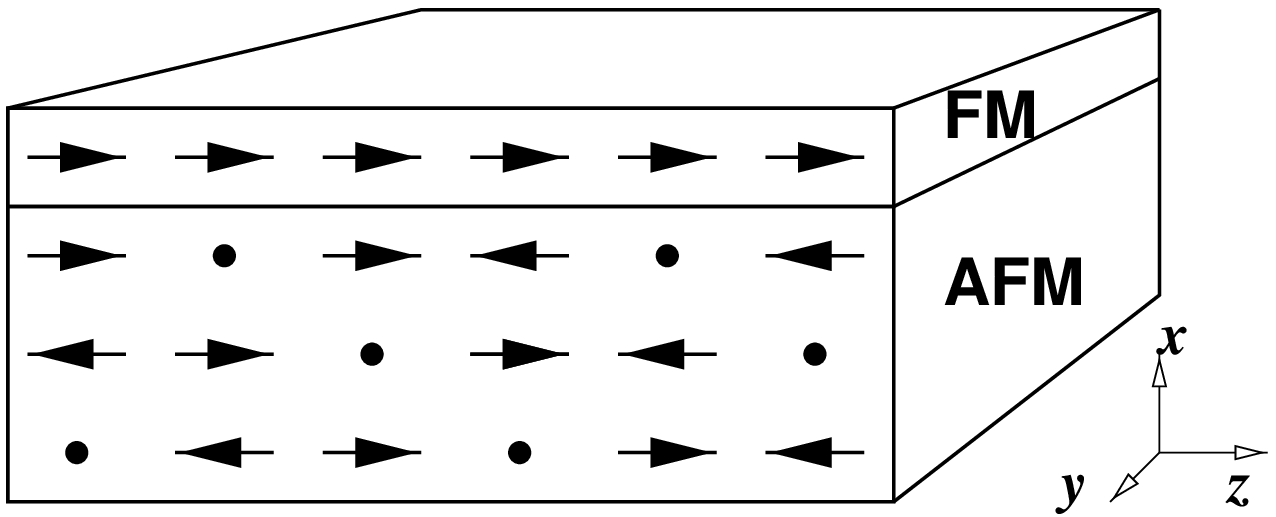}
  \caption{Sketch of the model with one FM layer and three diluted AFM 
    layers. The dots mark defects (non-magnetic ions).}
  \label{f:skizze}
\end{figure}
\fi

The system is described by a classical Heisenberg model,

\begin{eqnarray}
  {\cal H} = & \! \! \!- J_{\mathrm{FM}} \sum\limits_{\langle i, j \rangle}
                 {\ul S}_i \! \cdot \! {\ul S}_j - \sum\limits_i
                 \left( d_z S_{iz}^2 \! + \! d_x S_{ix}^2
                 \! + \! {\ul S}_i \! \cdot \! {\ul B} \right)  \nonumber \\
            & \! \! \! - J_{\mathrm{AFM}} \sum\limits_{\langle i, j \rangle}
                \epsilon_i \epsilon_j {\ul \sigma}_i \! \cdot \! {\ul \sigma}_j
                 -\sum\limits_i
                 \left( k_z \epsilon_i \sigma_{iz}^2 + \epsilon_i {\ul
                 \sigma}_i \! \cdot \! \ul B \right) \nonumber \\ 
            & - J_{\mathrm{INT}} \sum\limits_{\langle i, j \rangle}
                          \epsilon_i {\ul \sigma}_i \cdot {\ul S}_j, \nonumber
     \label{e:ham}                                     
\end{eqnarray} 
where the first line contains the energy contributions of the FM. Here,
the first term is the ferromagnetic nearest neighbor interaction with
exchange constant $J_{\mathrm{FM}}$.  The second term introduces an
easy axis in the FM ($z$-axis, anisotropy constant $d_z =
0.1J_{\mathrm{FM}}$) which sets the Stoner-Wohlfarth limit of the
coercive field, i. e. the zero temperature limit for magnetization
reversal by coherent rotation ($B_c = 2 d_z$, in our units, for a
field parallel to the easy axis).  The shape anisotropy is
approximated by the next term (anisotropy constant $d_x = -
0.1J_{\mathrm{FM}}$) leading to a magnetization which is
preferentially in the $y-z$-plane.  We checked, however, that its
value does not influence our results. The last term of the first line
is the Zeemann energy.

The second line describes the diluted AFM ($\epsilon_i = 0,1$;
dilution $p$) correspondingly except of the shape anisotropy.  For the
exchange constant of the AFM which mainly determines its N\'eel
temperature (also depending on the dilution and the uniaxial
anisotropy $k_z$) we set $J_{\mathrm{AFM}} = - J_{\mathrm{FM}}/2$.
Finally, the third line includes the interface coupling between FM and
AFM and for simplicity we assume $J_{\mathrm{INT}} = -
J_{\mathrm{AFM}}$).

Our magnetic field $B$ will always be along the $z$ axis.  In earlier
publications \cite{miltenyiPRL00,nowakJAP01,nowakPRB01} the AFM was
described by an Ising model. In the present work, we relax this
restriction on the AFM. In order to investigate a broader class of
systems for the AFM we vary the uniaxial anisotropy $k_z$ of the AFM.

\section{Results from Simulations} 
\label{s:results}

We use Monte Carlo methods with a heat-bath algorithm and single-spin
flip methods for the simulation of the model explained above. Each
spin is subject to a trial step consisting of a small deviation from
the original direction followed by a second trial step in form of a
total flip.  This two-fold trial step can take care of a broad range
of anisotropies starting from very soft spins up to the Ising limit
\cite{nowakARCP00}. We perform typically 25000 Monte Carlo steps per
spin for a complete hysteresis loop.

Since we are not interested in the critical behavior of the model
above, we do not perform a systematic finite-size analysis.  However,
in order to observe the domain structure of the AFM we have to
guarantee that typical length scales of the domain structure fit into
our system.  Therefore, we show here only results for rather large
systems of lateral extension $128 \times 128$. Nevertheless, we also
varied the system size and checked that our results are not influenced
by the system size as long as the system is not much smaller.

In our simulations the system is cooled from above to below the
ordering temperature of the AFM. During cooling the FM is long-range
ordered along the easy $z$ axis and its magnetization is practically
constant, resulting in a nearly constant exchange field for the AFM
monolayer at the interface. In addition to this exchange field the
external, magnetic field acts also on the whole AFM.

\ifodd\a
\begin{figure}[h]
  \epsfxsize=78mm \epsffile{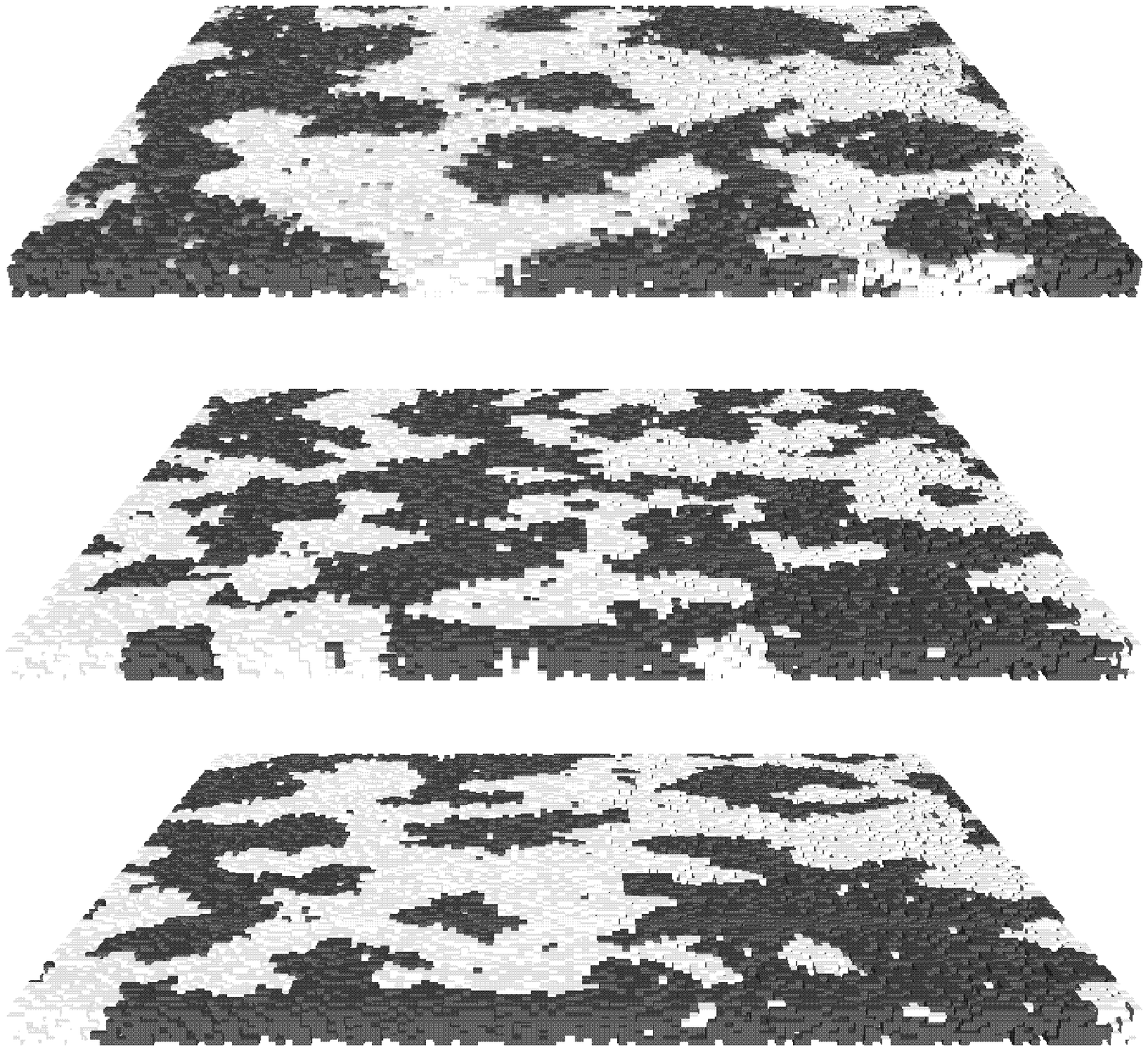} 
  \caption{Frozen domain states of a 40\% diluted AFM consisting of 6
  monolayers for different values of the AFM anisotropy, $k_{z} =0.1
  J_{\mathrm{FM}}, 1.0J_{\mathrm{FM}}, 30J_{\mathrm{FM}}$ (from
  top). The shading codes the $z$-component of the staggered
  magnetization. } \label{f:domains}
\end{figure}
\fi

As already argued in the section before, during the cooling procedure
the AFM becomes frozen in a domain state, the structure of which
depends on the system parameters. The influence on dilution
\cite{nowakPRB01} and the influence of the AFM film thickness
\cite{nowakJAP01} was already discussed before for the case of an
Ising AFM. In the present case, typical staggered domain
configurations of the bulk AFM are shown for three different values of
the AFM anisotropy (Fig.\ \ref{f:domains}). For low anisotropies $k_z
< J_{\mathrm{AFM}}$ domain walls have a width of the order of
$\sqrt{J_{\mathrm{AFM}} / k_z}$.  Even for the lowest anisotropy shown
in Fig.\ \ref{f:domains} the width is only of the order of a few
lattice constants which can hardly be detected in our figure. Also,
due to the dilution walls tend to follow the holes so that the wall
width is further reduced at those places.  Interestingly, the domain
structure itself depends also on $k_z$.  The system has the smallest
domains for an intermediate value of $k_z = J_{\mathrm{FM}}$ and not
for the Ising case corresponding to the high anisotropy limit $k_z =
30 J_{\mathrm{FM}}$ as one might expect. We will discuss the results
following from this behavior later in connection with the anisotropy
dependence of the EB.

Typical hysteresis loops taken after cooling in a field of $B =
0.25J_{\mathrm{FM}}$ are depicted in Fig.\ \ref{f:hysteresis}.  Shown
are results for the magnetization of the FM (upper figure) as well as
that of the AFM interface layer (lower figure). An exchange bias is
observed clearly and we determine the corresponding exchange bias
field as $B_{\mathrm{EB}} = (B^+ + B^-)/2$ where $B^+$ and $B^-$ are
those fields of the hysteresis loop branches for increasing and
decreasing field, where the easy axis component of the magnetization
of the FM becomes zero.

\ifodd\a
\begin{figure}[h]
  \epsfxsize=70mm \epsffile{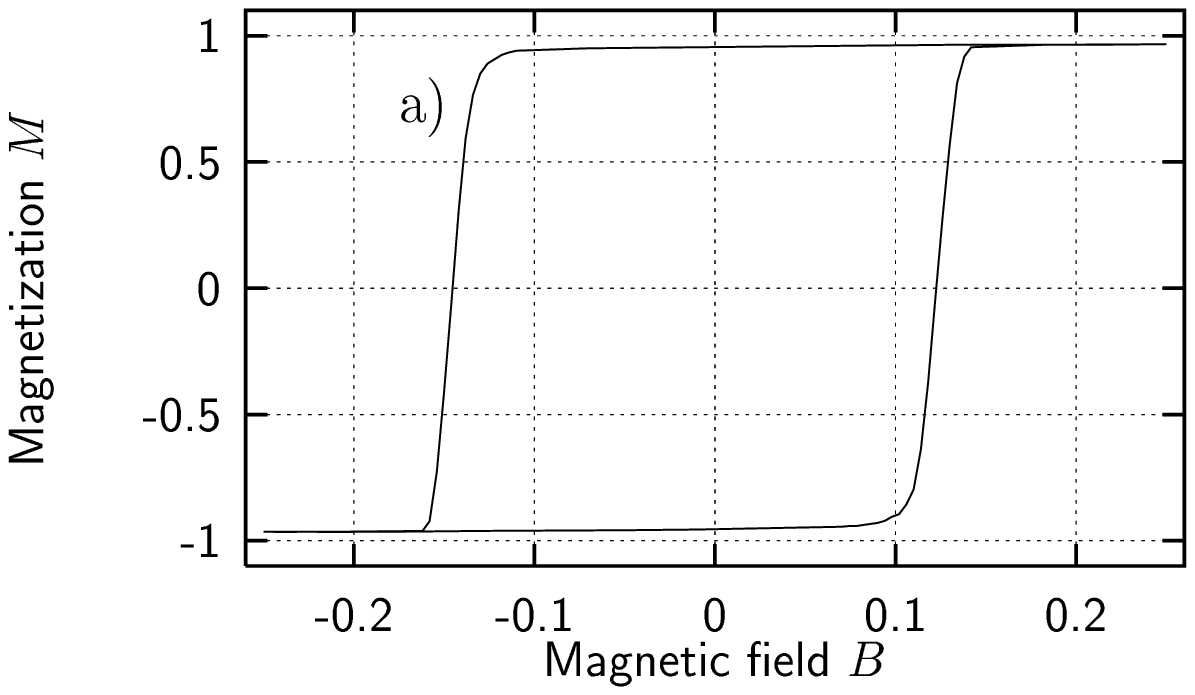} \epsfxsize=70mm
  \epsffile{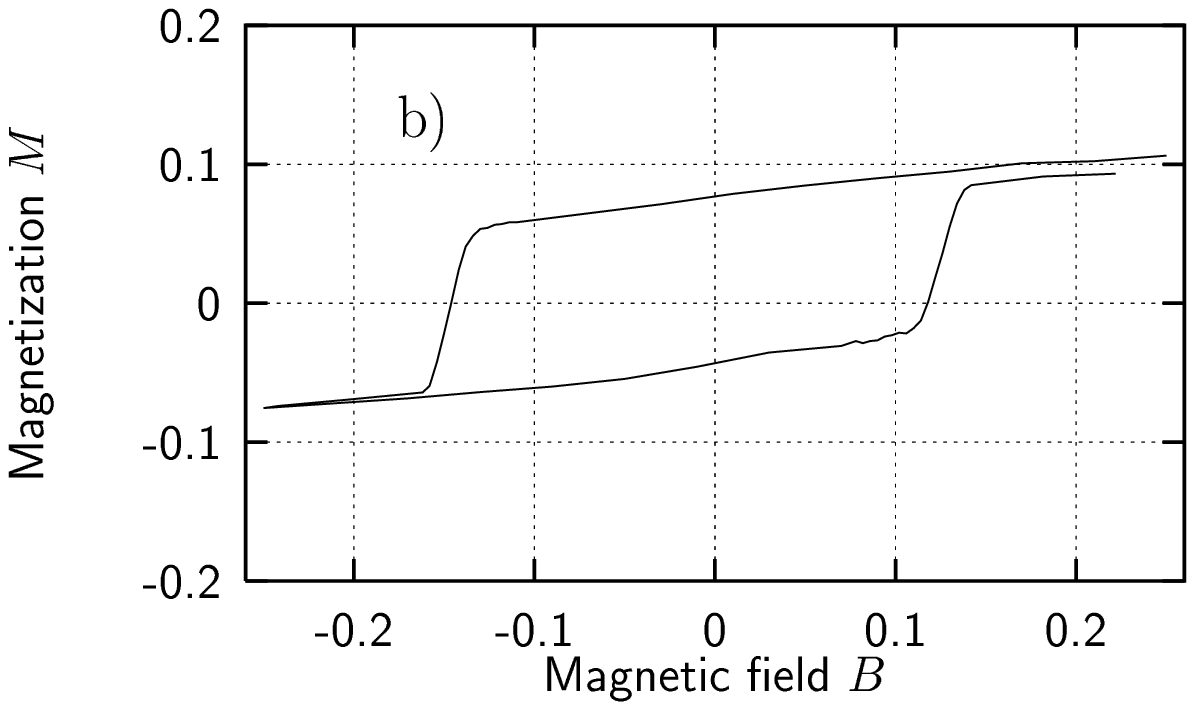}
  \caption{Typical hysteresis loop along $z$ of a) the FM and b) the
  interface layer of the AFM, for a dilution $p=0.4$ and an AFM
  thickness $t=2$. The magnetization is in units of the saturation
  value and the field in units of $J_{\mathrm{FM}}$.}
  \label{f:hysteresis}
\end{figure}
\fi

The interface magnetization of the AFM also shows a hysteresis,
following the coupling to the FM. Additionally, its curve is shifted
upwards due to the fact that after field cooling the AFM is in a
domain state with a surplus magnetization. The upward shift of the
hysteresis loop for the interface AFM proves the existence of remanent
magnetization in the AFM domains. This shifted interface magnetization
of the AFM acts as an additional effective field on the FM, resulting
in EB.  The magnitude of the EB field strongly depends on the amount
of this upward shift. Note that the shift is of the order of a few
percent of the saturation magnetization of the AFM while approximately
10\% of the spins of the AFM contribute to the AFM hysteresis. The
saturation field for the AFM is much larger than that of the FM so
that the AFM is never saturated during the simulation.

In absence of any anisotropy in the FM and at very low dilution of the
AFM we observe a perpendicular coupling between FM and AFM.  The
magnetization reversal in the FM is here by coherent rotation. The
picture changes with increasing uniaxial anisotropy in the FM and upon
further dilution of the AFM. The magnetization reversal in the FM is
now by domain wall motion and the perpendicular coupling becomes less
significant.  This is because uniaxial anisotropies in both the FM and
the AFM having the same axis no longer lead to an energy minimum for a
perpendicular coupling across the interface. Moreover AFM spins with
missing AFM neighbors can lower their energy by rotating parallel to
their FM neighbors. Therefore, in the framework of our calculations a
spin-flop coupling is not an essential mechanism for EB.

\ifodd\a
\begin{figure}[h]
  \epsfxsize=60mm
  \epsffile{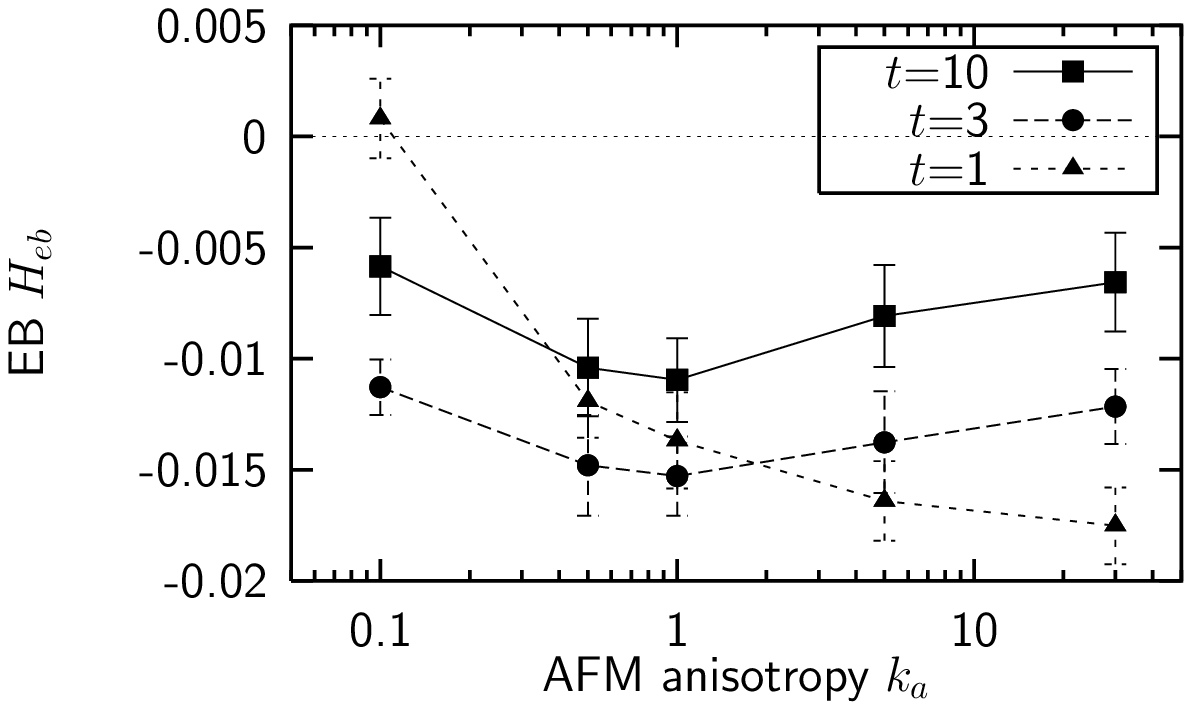}
  \caption{EB field versus anisotropy of the AFM for
    different AFM thicknesses (numbers of AFM layers). Field and
    anisotropy are in units of $J_{\mathrm{FM}}$.}
 \label{f:anisotropy}
\end{figure}
\fi

We have calculated the EB field for a wide range of values of $k_{z}$,
starting from very soft spins to rigid, Ising-like spins. Fig.
\ref{f:anisotropy} shows result for different thicknesses of the AFM
and for a dilution of $p=0.4$. Interestingly, we find a peak in the EB
field at an intermediate value of $k_{z}$ for a sufficiently thick AFM
while at lower thicknesses the EB field increases with the anisotropy
and saturates in the Ising limit.

The key for the understanding of EB is the understanding of AFM domain
configurations and domain walls. AFM domains are required to carry a
surplus magnetization at the interface which must be stable along the
$z$ direction during hysteresis in order to produce any EB.  In
general, one might expect that the most stable domain configurations
are obtained for the Ising limit. But Fig.\ \ref{f:anisotropy} and
also Fig.\ \ref{f:domains} suggest that the behavior of domains is
more complex. Let us start considering the Ising limit where some
domain wall is formed upon field cooling. This domain wall
preferentially passes through defects thereby minimizing the exchange
energy and at the same time it gathers magnetization thereby lowering
the Zeeman energy.  When the anisotropy $k_{z}$ is decreased the
energy to create a domain wall will decrease. Thus the system will
respond by roughening the domain boundaries (see Fig.\ 
\ref{f:domains}, where the domain configuration in the middle is more
complex than the lower one which represents the Ising limit). This
roughening enhances the possibility for the domains to carry any
surplus magnetization and hence the EB will increase.  However, there
exists a counter effect.  While further decreasing $k_{z}$ the width
of the domain wall increases, so that less energy can be saved through
the dilution.  Hence, for still lower anisotropy the domain walls will
smoothen thereby lowering again the exchange energy (see once again
Fig.\ \ref{f:domains}, now comparing the domain configuration in the
middle with the upper one for still lower anisotropy which has much
smoother domain walls). Since flat walls carry less remanent
magnetization the EB will decrease now with decreasing anisotropy.
The compromise between these two opposite effects is achieved at some
intermediate value of $k_{z}$ where the bias shows a peak.

However, the peak disappears at lower values of $t$. This happens
since for only one monolayer of AFM we are close to the percolation
threshold where the domain walls pass nearly exclusively through the
defects costing very little or no energy.  Therefore the first
mechanism discussed above is less important and the EB increases with
$k_{z}$ till it saturates in the Ising limit.

\section{Conclusions}
In conclusion, we find that the domain state model for EB proposed
originally for the Ising AFM is not restricted to this limit. Rather
under certain combination of thickness and dilution of the AFM, the
softness of the AFM spins can lead to an even stronger bias field.
Since disorder in the AFM of an exchange bias system is rather common,
our model yields a general understanding of the microscopic origin of
exchange bias.  Within our model there are several properties which
influence the bias field, such as dilution, thickness and anisotropy
of the AFM. Although a qualitative understanding regarding the
dependence of EB on these parameters has been achieved, a quantitative
study of the domain structure both at the interface and in the bulk of
the AFM would provide a deeper understanding to the problem.

\section*{Acknowledgments}
This work has been supported by the Deutsche Forschungsgemeinschaft
through SFB 491 and Graduiertenkolleg 277.

\ifodd\a
\end{document}
\fi

\begin{figure}[p]
  \caption{Sketch of our model with one FM layer and three AFM
  layers.}
  \label{f:skizze}
\end{figure}

\begin{figure}[p]
  \caption{Frozen domain states of a 40\% diluted AFM consisting of 6
    monolayers for different values of the AFM anisotropy, $k_{z}
    =0.1 J_{\mathrm{FM}}, 1.0J_{\mathrm{FM}},
      30J_{\mathrm{FM}}$ (from top).}
 \label{f:domains}
\end{figure}

\begin{figure}[p]
  \caption{Typical hysteresis loop along $z$ of (a) the FM and (b)
    the interface monolayer of AFM, for $p=0.4$ $t=2$. The net
    magnetization is shown in units of the saturation magnetization  
    and the field in units of $J_{\mathrm{FM}}$.} 
  \label{f:hysteresis}
\end{figure}

\begin{figure}[p]
  \caption{Exchange bias field versus anisotropy of the AFM for
  different AFM thicknesses (numbers of AFM layers).}
 \label{f:anisotropy}
\end{figure}

\clearpage
\setcounter{figure}{0}
\begin{figure}[h]
  \epsfxsize=60mm
  \epsffile{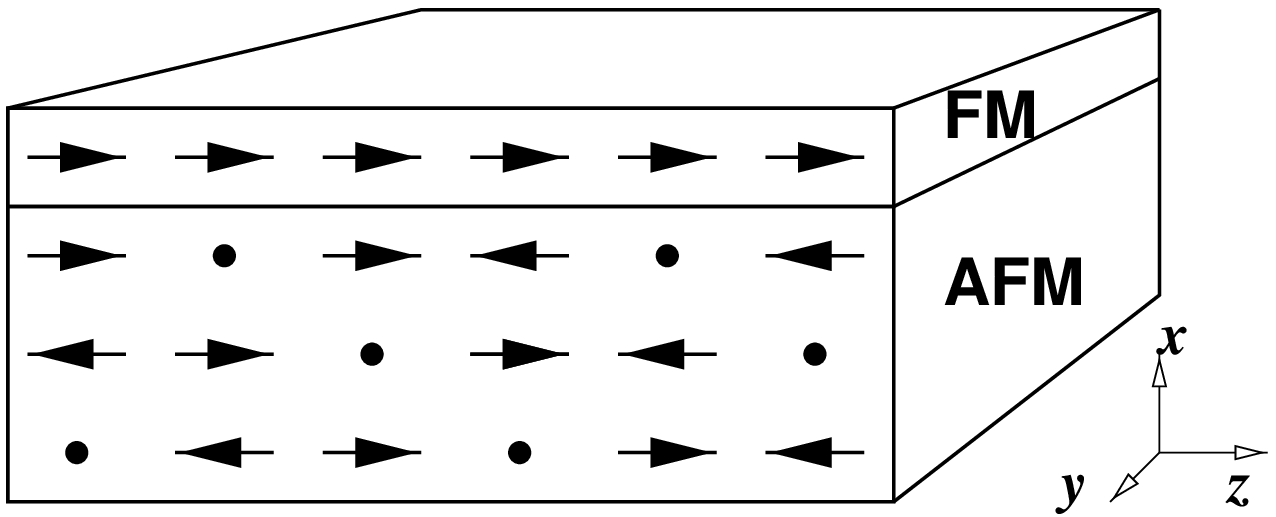}
  \vspace{3cm}
  \caption{Nowak et al.}
\end{figure}

\newpage
\begin{figure}[h]
  \epsfxsize=70mm
  \epsffile{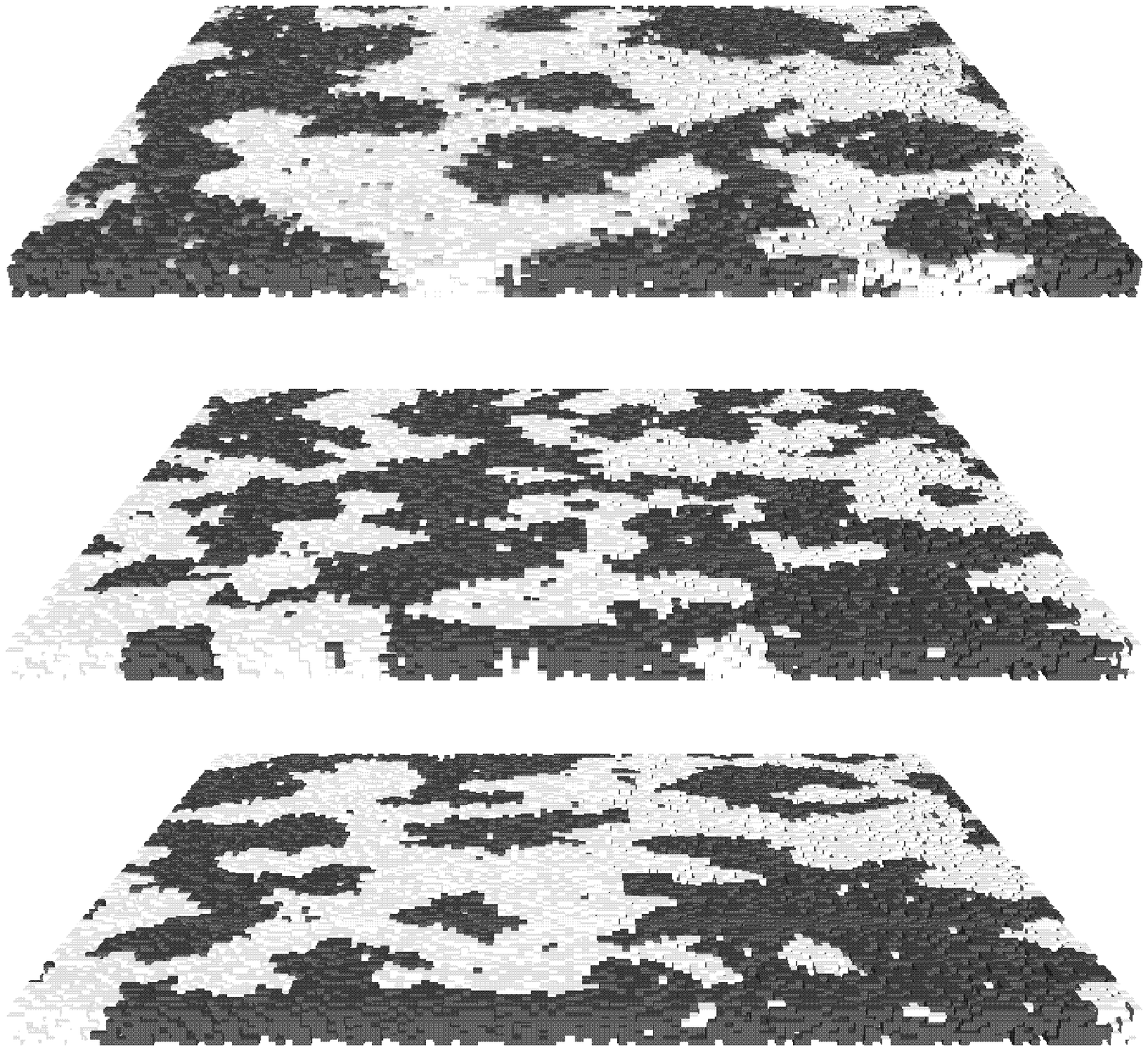}
  \vspace{3cm}
  \caption{Nowak et al.}
\end{figure}

\newpage
\begin{figure}[h]
  \epsfxsize=70mm  \epsffile{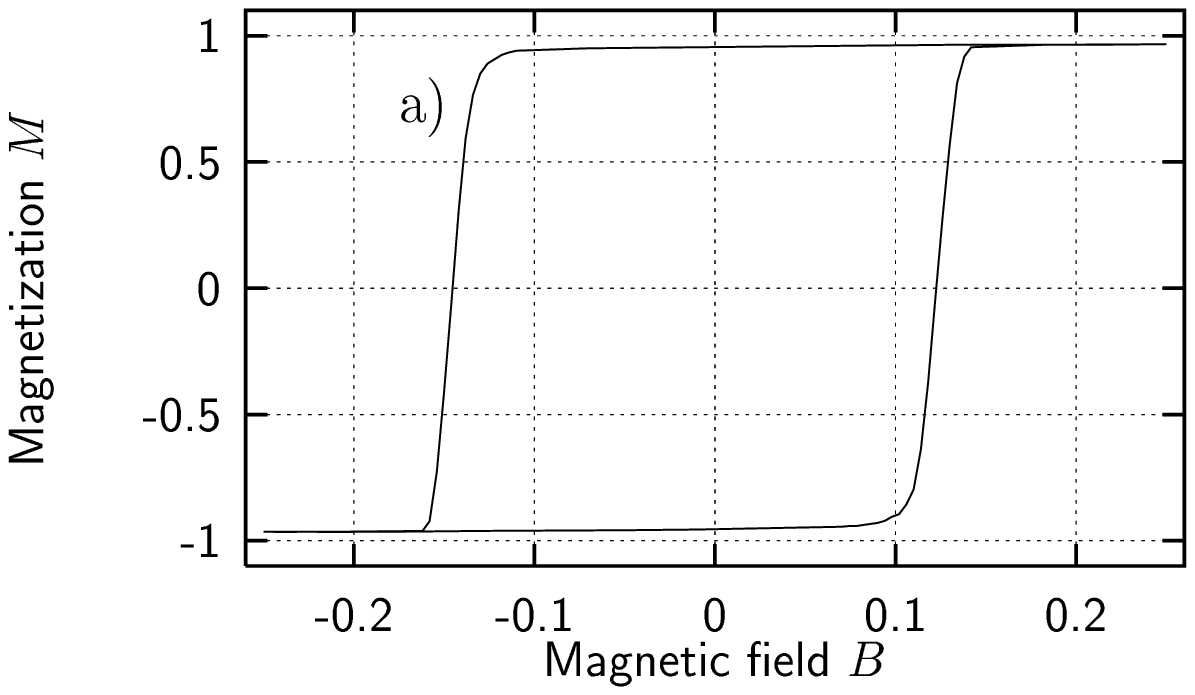}
  \epsfxsize=70mm  \epsffile{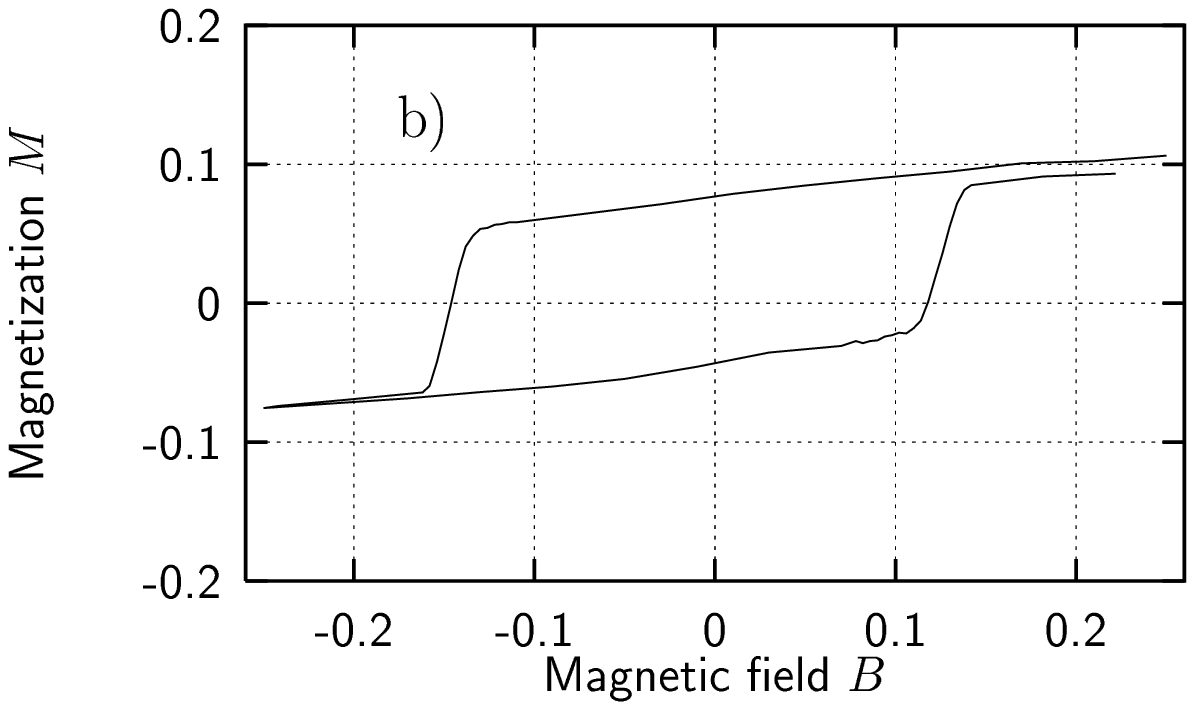}
  \vspace{3cm}
  \caption{Nowak et al.}
\end{figure}

\newpage
\begin{figure}[h]
  \epsfxsize=70mm
  \epsffile{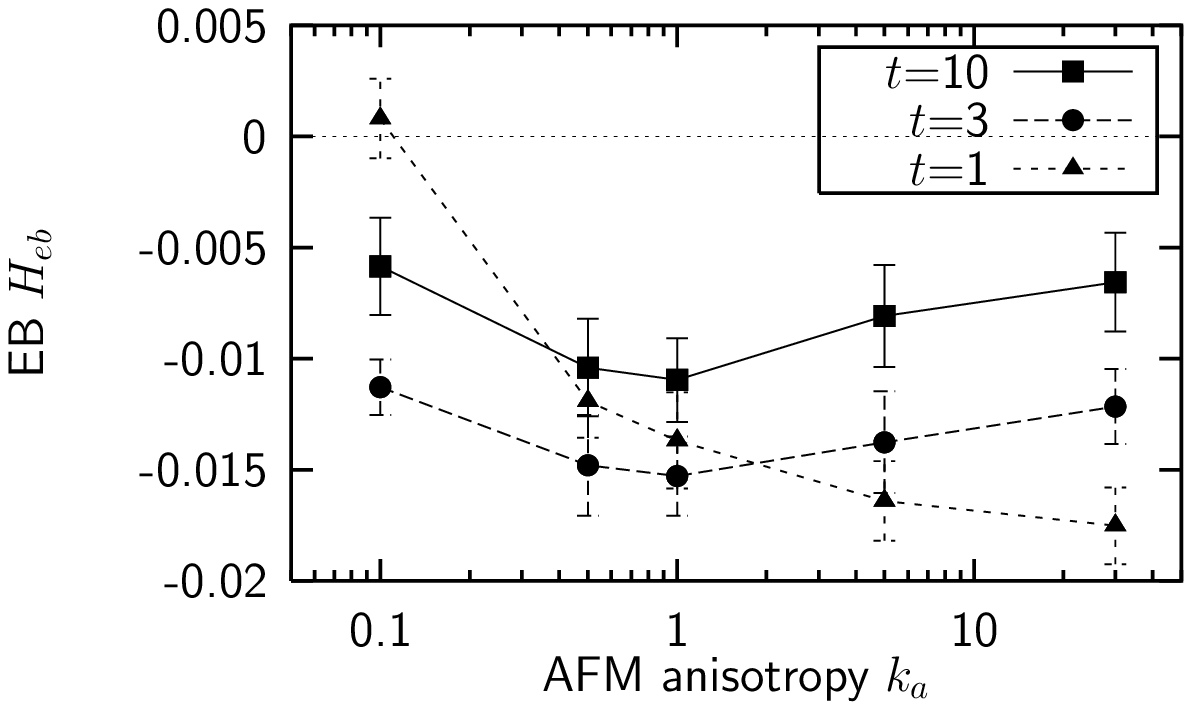}
  \vspace{3cm}
  \caption{Nowak et al.}
\end{figure}


\begin{thebibliography}{10}

\bibitem{meiklejohnPR56}
W.~H. Meiklejohn and C.~P. Bean, Phys.\ Rev. {\bf 102},  1413  (1956).

\bibitem{noguesJMMM99}
J. Nogu\'es and I.~K. Schuller, J.\ Magn.\ Magn.\ Mat. {\bf 192},  203  (1999).

\bibitem{malozemoffPRB87}
A.~P. Malozemoff, Phys.\ Rev.\ B {\bf 35},  3679  (1987).

\bibitem{koonPRL97}
N.~C. Koon, Phys.\ Rev.\ Lett. {\bf 78},  4516  (1998).

\bibitem{schulthessPRL98}
T.~C. Schulthess and W.~H. Butler, Phys.\ Rev.\ Lett. {\bf 81},  4516  (1998).

\bibitem{schulthessJAP99}
T.~C. Schulthess and W.~H. Butler, J.\ Appl.\ Phys. {\bf 85},  5510  (1999).

\bibitem{miltenyiPRL00}
P. Milt\'enyi, M. Gierlings, J. Keller, B. Beschoten, G. G\"untherodt, U.
  Nowak, and K.~D. Usadel, Phys.\ Rev.\ Lett. {\bf 84},  4224  (2000).

\bibitem{mouginPRB01}
A. Mougin, T. Mewes, M. Jung, D. Engel, A. Ehresmann, H. Schmoranzer, J.
  Fassbender, and B. Hillebrands, Phys.\ Rev.\ B {\bf 63},  60409  (2001).

\bibitem{nowakJAP01}
U. Nowak, A. Misra, and K.~D. Usadel, J.\ Appl.\ Phys. {\bf 89},  7269  (2001).

\bibitem{nowakPRB01}
U. Nowak and K.~D. Usadel, Phys.\ Rev.\ B  (2001), in preparation.

\bibitem{kleemannIJMP93}
W. Kleemann, Int.\ J.\ Mod.\ Phys.\ B {\bf 7},  2469  (1993).

\bibitem{belangerBOOK98}
D.~P. Belanger,  in {\em Spin Glasses and Random Fields}, edited by A.~P. Young
  (World Scientific, Singapore, 1998).

\bibitem{fishmanJPC79}
S. Fishman and A. Aharony, J.\ Phys.\ C {\bf 12},  L729  (1979).

\bibitem{cardyPRB84}
J.~L. Cardy, Phys.\ Rev.\ B {\bf 29},  505  (1984).

\bibitem{imbriePRL84}
J.~Z. Imbrie, Phys.\ Rev.\ Lett. {\bf 53},  1747  (1984).

\bibitem{bricmontPRL87}
J. Bricmont and A. Kupiainen, Phys.\ Rev.\ Lett. {\bf 59},  1829  (1987).

\bibitem{montenegroPRB91}
F.~C. Montenegro, A.~R. King, V. Jaccarino, S.-J. Han, and D.~P. Belanger,
  Phys.\ Rev.\ B {\bf 44},  2255  (1991).

\bibitem{nowakPRB91}
U. Nowak and K.~D. Usadel, Phys.\ Rev.\ B {\bf 44},  7426  (1991).

\bibitem{imryPRL75}
Y. Imry and S. Ma, Phys.\ Rev.\ Lett. {\bf 35},  1399  (1975).

\bibitem{nowakPRB92}
U. Nowak and K.~D. Usadel, Phys.\ Rev.\ B {\bf 46},  8329  (1992).

\bibitem{esserPRB97}
J. Esser, U. Nowak, and K.~D. Usadel, Phys.\ Rev.\ B {\bf 55},  5866  (1997).

\bibitem{birgeneauJSP84}
R.~J. Birgeneau, R.~A. Cowley, G. Shirane, and H. Yoshizawa, J.\ Stat.\ Phys.
  {\bf 34},  817  (1984).

\bibitem{belangerJAP85}
D.~P. Belanger, M. Rezende, A.~R. King, and V. Jaccarino, J.\ Appl.\ Phys. {\bf
  57},  3294  (1985).

\bibitem{villainPRL84}
J. Villain, Phys.\ Rev.\ Lett. {\bf 52},  1543  (1984).

\bibitem{hanPRB92}
S.-J. Han, D.~P. Belanger, W. Kleemann, and U. Nowak, Phys.\ Rev.\ B {\bf 45},
  9728  (1992).

\bibitem{nowakARCP00}
U. Nowak,  in {\em Annual Reviews of Computational Physics IX}, edited by D.
  Stauffer (World Scientific, Singapore, 2000), p.\ 105.

\end{thebibliography}
\end{document}